\documentclass[%
 reprint,
 amsmath,amssymb,
 aps,
]{revtex4-2}

\usepackage{graphicx}
\usepackage{dcolumn}
\usepackage{bm}

\begin{document}

\preprint{APS/123-QED}

\title{Optical contrast-based determination of number of layers for two-dimensional van der Waals magnet Fe$_3$GeTe$_2$}
\thanks{A footnote to the article title}%

\author{Neesha Yadav, Sandeep, Pintu Das}
\affiliation{Department of Physics, Indian Institute of Technology, Delhi, Hauz Khas, New Delhi, India}
 \altaffiliation[]{Department of Physics, Indian Institute of Technology, Delhi, Hauz Khas, New Delhi, India}




\date{\today}

\begin{abstract}
Recent advances in revealing intrinsic magnetism in two-dimensional (2D) materials have highlighted their potential for future spintronic applications, driven by their novel physical properties, promising for future spintronic devices. In order to explore layer dependent magnetic behavior, in general, mechanically exfoliated flakes from high-quality single crystals are used. It is crucial to determine the number of layers of these materials accurately. In the absence of an efficient and quick method, researchers often rely on atomic force microscopy (AFM) imaging to identify their number of layers. In this work, we report an optical contrast study as a quick and cost-effective technique to determine the number of layers of Fe$_3$GeTe$_2$ (FGT). Here, we observed a linear relationship between the optical contrast (derived from optical microscopic images) observed for mechanically exfoliated FGT nano-flakes and their thickness, as measured by the AFM imaging method. This technique requires no additional equipment; it relies solely on a conventional optical microscope. Additionally, our results reveal a thickness-dependent evolution of the intensity; in contrast, the Raman frequency demonstrates no significant dependence on layer thickness. Also, our studies reveal two additional Raman modes of FGT, at the frequency of 129\,cm$^-1$ \& 190\,cm$^-1$. Both modes show the intensity dependence on the thickness of FGT, same as out-of-plane (A$_{1g}$) Raman modes. 
\end{abstract}

\maketitle


\section{Introduction}
Two-dimensional (2D) van der Waals magnetic materials offers an exciting platform for exploring magnetism down to monolayer thickness, making them potential candidates for future spintronic devices \cite{a3,a2}. Among all 2D magnets, the Fe-based materials possess a high ferromagnetic transition temperature ($T_\mathrm{C}$), which enables the fabrication of near-room-temperature devices for practical applications. Here, we study the Fe$_3$GeTe$_2$ (FGT), comprising the ($T_\mathrm{C}$) around 200\,K. FGT exhibits long-range ferromagnetic order down to atomic thickness, as well as metallic behavior, which enables the exploration of its electrical transport properties \cite{a5,a4,a3}. The mechanical exfoliation of the crystal provides high-quality and low-cost fabrication of samples. FGT is air sensitive and oxidizes in the ambient environment, which can alter its ferromagnetic properties by introducing an oxide layer at the interface. The oxide layer gives rise to the antiferromagnetic signature in the system. The thin, exfoliated samples of FGT are more susceptible to oxidation compared to bulk crystals; therefore, fabrication needs to be carried out inside an inert gas-filled glove box. The electrical and magnetic properties of 2D magnets are highly sensitive to their thickness or the number of layers \cite{a6}. Therefore, accurately determining the number of layers in 2D magnetic materials is crucial for reliably studying their electrical and magnetic behavior, such as in the case of FGT. Conventionally, atomic force microscopy (AFM) is used to determine the thickness of samples; however, it is challenging to use inside a glove box, as it requires additional equipment to characterize the thickness, as well as the necessity to remove the sample from the glove box for thickness characterization. Extracting the sample affects the quality of the material, which plays a crucial role in the magnetic behavior of FGT. Therefore, an efficient method is required to determine the number of layers of FGT that can be employed inside the glove box. Optical contrast analysis is a well-established technique for thickness characterization of 2D materials, including graphene and MoS$_2$ \cite{a14,a15,a16}; nevertheless, its use for 2D magnetic materials has not yet been systematically investigated. In this report, we introduce an efficient and non-destructive method that utilizes the optical contrast of mechanically exfoliated nanoflakes to identify the number of layers of FGT. This method provides in situ detection of the number of layers without compromising the quality of the FGT material. Moreover, the Raman spectroscopy is investigated on the same nanoflakes to complement the results. Additionally, we report two additional Raman modes of FGT at frequencies of 129\,cm$^-1$ \& 190\,cm$^-1$ which further need to be studied.
\section{Methods}
\subsection{Fabrication of 2D nanoflakes}
The 2D nanoflakes of FGT are mechanically exfoliated on the Si/SiO$_2$ substrate using the conventional heat-assisted scotch tape method. Initially, the Si/Sio$_2$ substrate is cleaned ultrasonically using acetone, isopropyl alcohol (IPA), and deionized (DI) water, respectively, followed by a nitrogen blow dry. The FGT is thinned down by the multiple peelings of the scotch tape from the parent single-crystal, procured from HQ Graphene, and transferred onto the Si/SiO$_2$ substrate. Before transferring the flakes to the substrate, the substrate is heated to remove the air traps or moisture to achieve good adhesion of the flakes on the substrate.
\subsection{Characterization of the number of layers of fabricated 2D nanoflakes}
The prepared nanoflakes are microscopically inspected under the "Leica DM 750 M" optical microscope in the reflective mode. The optical contrast of the samples is determined using the ImageJ software. The thicknesses of the nanoflakes are characterized by the Asylum Research atomic force microscopy (AFM) in tapping mode. The scan rate of 0.3 Hz is used for the AFM measurements, and the freely accessible WSXM software is used to analyze the recorded data. Raman spectroscopy is performed by using the Renishaw inVia Raman microscope. The 20 mW laser power is exposed for 30 seconds with three accumulations to perform the experiments by using the laser of wavelength, 532\,nm.
\section{Results \& Discussion}
\begin{figure}[h!]
    \centering
    \includegraphics[width=1\linewidth]{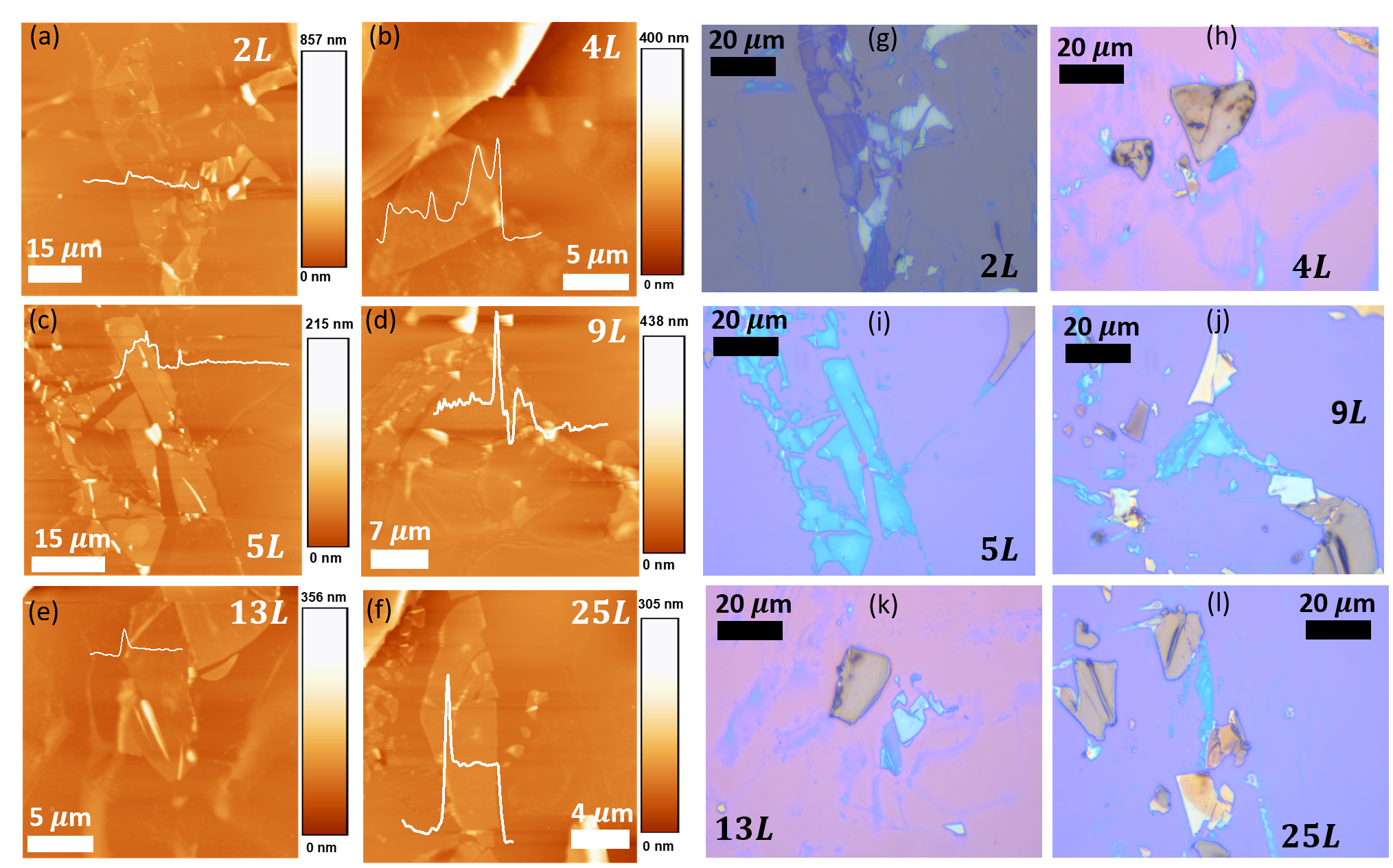}
    \caption{(a) - (f) AFM images of mechanically exfoliated nanoflakes with the thickness analysis. (g) - (l) The corresponding optical microscopic images of the flakes.}
    \label{Optical contrast}
\end{figure}

\begin{figure}[h!]
     \centering
     \includegraphics[width=1\linewidth]{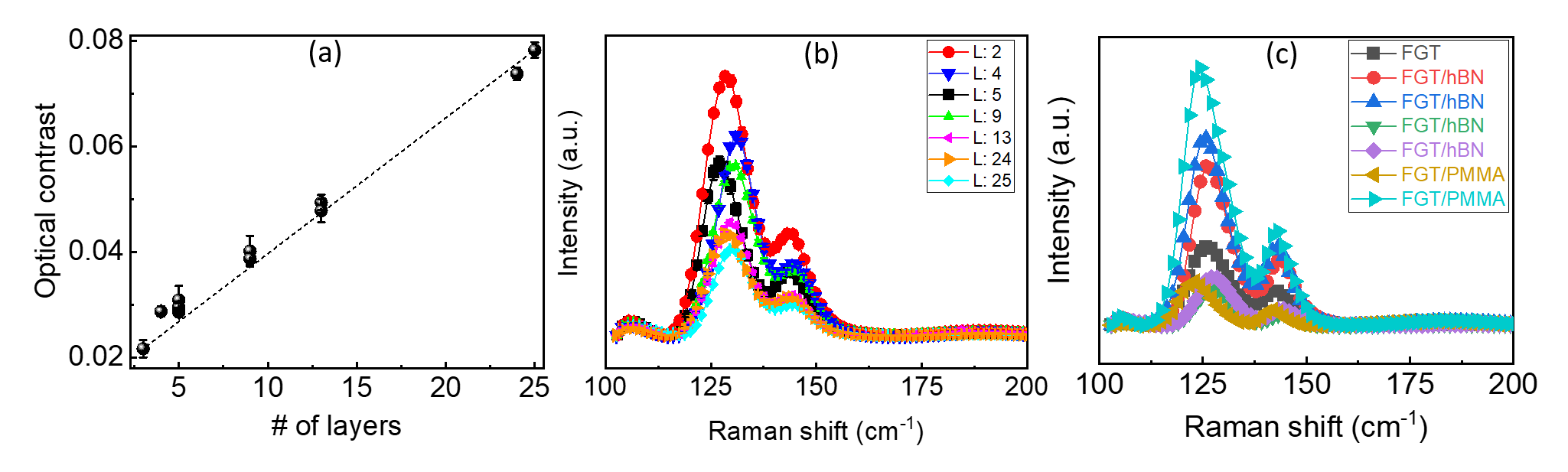}
     \caption{Optical contrast and Raman spectroscopy analysis of the FGT samples. (a) Thickness-dependent optical contrast of FGT exfoliated on Si/SiO$_2$ substrate. (b) The Raman spectra of several FGT samples with varying layer numbers from 2 to 25 monolayers. (c) Raman spectra of FGT, encapsulated with hBN and PMMA.}
     \label{Raman spectroscopy}
 \end{figure}

The initial thickness characterization of the FGT nanoflakes is accomplished using conventional AFM experiments. Figure \ref{Optical contrast} (a) - (f) shows the AFM images of the exfoliated flakes. The corresponding microscopic images of the flakes are shown in Figure \ref{Optical contrast} (g) - (l). The thickness of them is determined by the line profile of the AFM images. The line profile has been taken at the same position where the graph is plotted in the Figure. The determined thickness is further transformed into the number of layers of FGT by dividing the thickness of the FGT monolayer (0.8\,nm). Every AFM and microscopic image contains a characterized number of layers, as shown in Figure \ref{Optical contrast}.\\
 Further, we study the optical contrast of the same flakes in order to relate the thickness of FGT to optical contrast. The intensity of the substrate and flakes is determined using optical microscopic images of the flakes using ImageJ software. The Optical contrast is calculated by the formula given below -
\begin{center}
       Optical Contrast = $\frac{\mathrm{I}_s - \mathrm{I_f}}{\mathrm{I_s} + \mathrm{I_f}}$
    \end{center}
    where --\\
    I$_s$ = Intensity of substrate\\
    I$_f$ = Intensity of flake\\
 For more than 25 layers of FGT, the optical contrast is determined to be negative due to the higher reflection from the thick flakes than from the substrate. Therefore, the optical contrast method can be used up to 25 layers of FGT efficiently. The number of layers, identified by AFM, is plotted with calculated optical contrast in Figure \ref{Raman spectroscopy} (a), and the graph shows a linear relation between the optical contrast and the number of layers of FGT. The error bar represents the standard deviation of the optical contrast, calculated as the average of 10 repetitions of the intensity measurements for flakes and substrates. The optical contrast study is an efficient tool to determine the number of layers of FGT, as it shows the systematic thickness dependence.

 \begin{figure}[h!]
    \centering
    \includegraphics[width=1\linewidth]{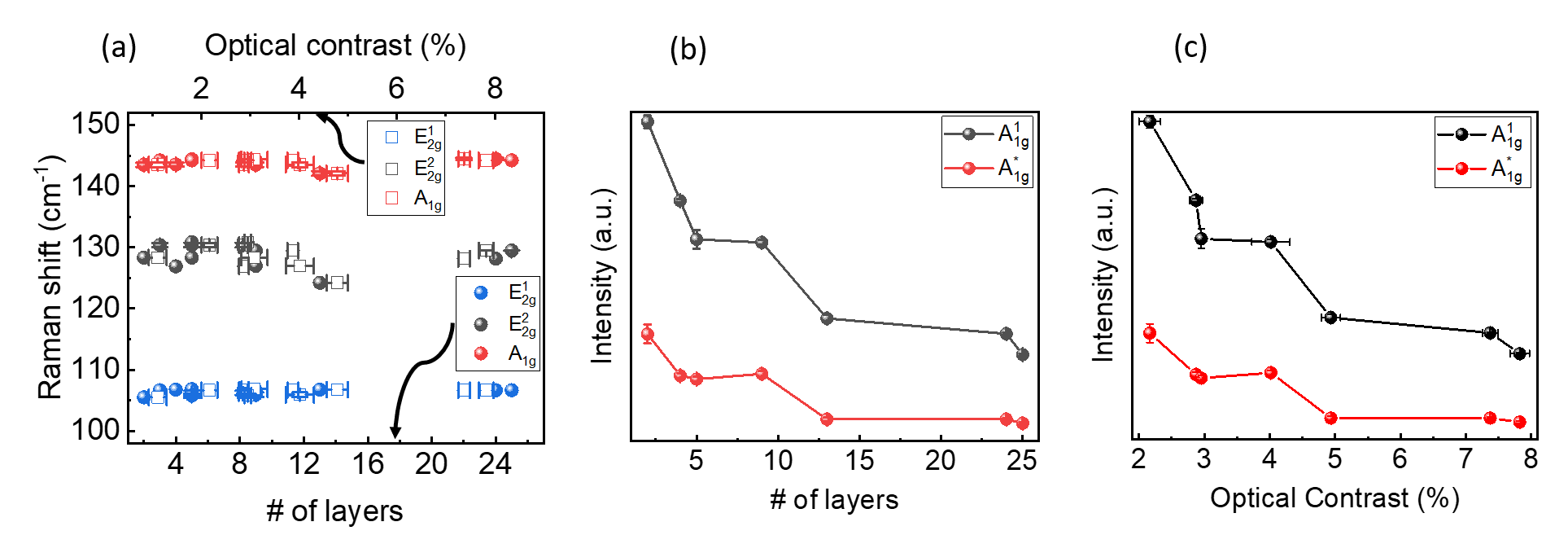}
    \caption{Raman spectroscopy analysis of various FGT samples of different thicknesses. (a)  The peak positions of all three peaks, $E_{2g}^1$, $A_{1g}^1$ \& $A_{1g}^*$ with the number of layers of FGT and optical contrast. The Raman intensity of $A_{1g}^1$ mode \& $A_{1g}^*$ mode of FGT with the number of layers (b) and optical contrast (c).}
    \label{Raman analysis}
\end{figure}
\begin{figure}
    \centering
    \includegraphics[width=1\linewidth]{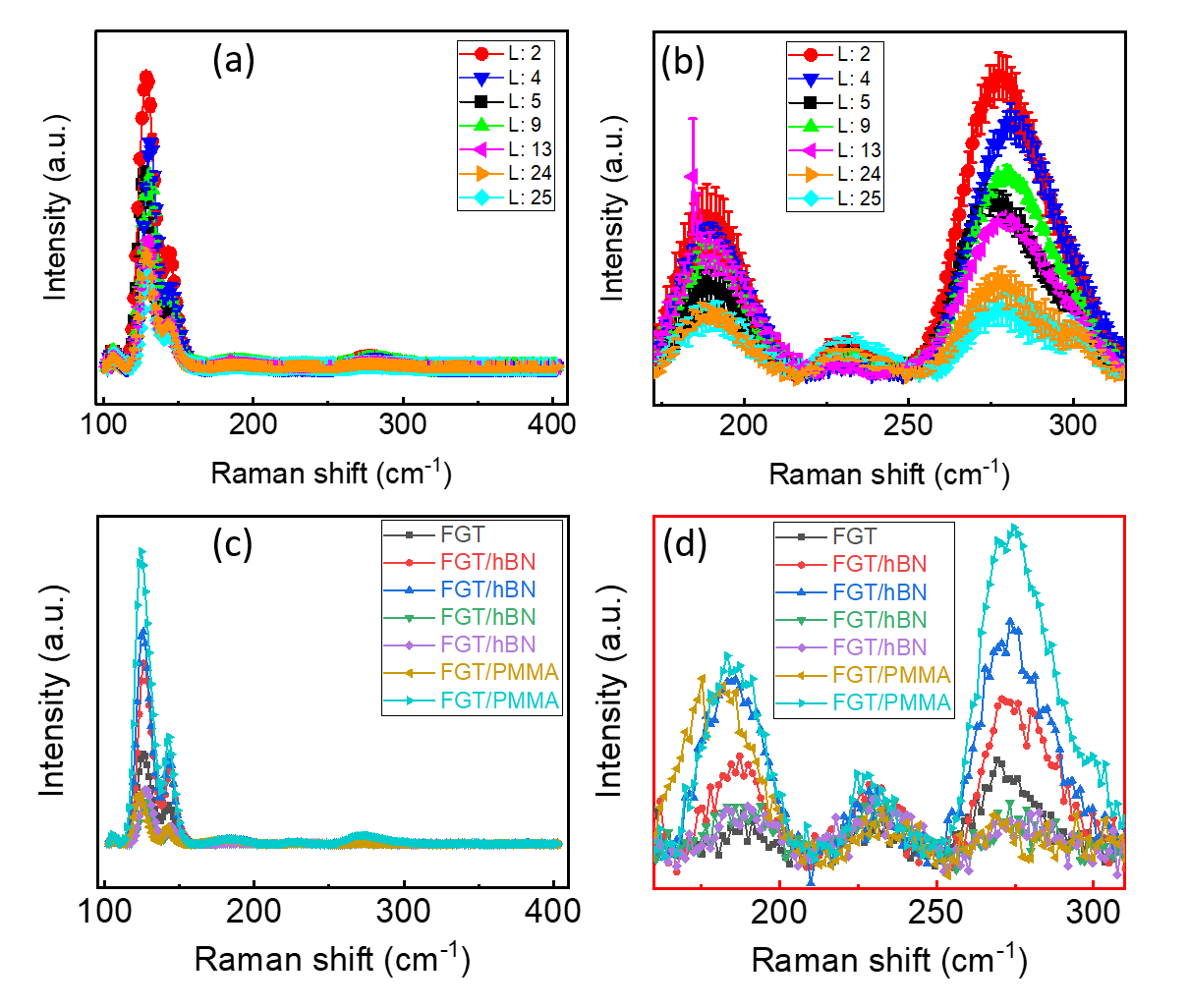}
    \caption{(a) \& (c) The Raman spectra of the FGT and encapsulated FGT nanoflakes at higher frequencies, respectively. (b) \& (d) The zoomed spectra between in the range of 150-325 cm$^{-1}$ as shown in (a) \& (c).}
    \label{Raman modes}
\end{figure}
 Furthermore, we carried out the Raman spectroscopy experiments on the same FGT nanoflakes. Raman spectroscopy is a non-destructive technique that can be used to determine the number of layers of 2D material due to their thickness-dependent Raman modes. Here, Raman spectroscopy is used to study numerous FGT nanoflakes and examine the vibrational levels of FGT, as shown in Figure \ref{Raman spectroscopy} (b). The Raman spectra of FGT show one weak and two prominent Raman active modes at lower frequencies; positioned at 105 cm$^{-1}$, 129 cm$^{-1}$ \& 145 cm$^{-1}$. In the Raman spectra of pristine FGT, two type of Raman active modes, E$_\mathrm{2g}$ \& A$_\mathrm{1g}$ exist which are due to the in-plane vibrations of Fe, Ge \& Te atoms and the out of plane vibrations of Te \& Fe atoms, respectively \cite{a1} but the Raman spectra of our FGT nanoflakes shows the significant discrepancies with the theoretically predicted Raman peaks of pristine FGT \cite{a1}. Apart from two Raman active modes at 105 cm$^{-1}$ (E$_\mathrm{2g}^1$), 145 cm$^{-1}$ (A$_\mathrm{1g}^1$), we observed an additional Raman-active mode of FGT, positioned at 129 cm$^{-1}$ (A$_\mathrm{1g}^*$), in our spectra. The possible reason behind the discrepancies in the spectra could be the oxidation of FGT, as FGT is exfoliated under ambient conditions. To confirm whether the additional mode occurs due to the oxidation of FGT, we exfoliated a few FGT flakes inside the Ar-filled glove box. The flakes are covered with hexagonal boron nitride (hBN) or polymethylmethacrylate (PMMA) before being removed from the Glove box to prevent oxidation of the FGT. The Raman spectra of these nanoflakes show the same three peaks at lower frequencies, as shown in Figure \ref{Raman spectroscopy} (c). Therefore, it is clear that the third additional peak is the intrinsic Raman mode of FGT, not due to the oxidation of FGT. Each Raman spectrum has been reproduced three times, and the standard deviation is used as the error bars.\\
 To investigate the thickness dependence of Raman modes, we studied the position of Raman modes with optical contrast and the number of layers, shown in Figure \ref{Raman analysis} (a). The positions of all three Raman-active modes do not exhibit a significant change with the number of layers. All three modes are positioned at almost the same frequency, irrespective of the FGT thicknesses. Therefore, the Raman shift is not an effective tool to determine the number of layers of FGT. However, the Raman intensity of both of the prominent Raman modes (A$_\mathrm{1g}^1$, \& A$_\mathrm{1g}^*$) decreases with the increase of the number of layers of FGT, shown in Figure \ref{Raman analysis} (b \& c). As the thickness of FGT increases, the stacking of the atoms in the c direction slows down the out-of-plane vibrations of the atoms, which consequently suppresses the intensities of the (A$_\mathrm{1g}$) Raman mode. As adding more layers increases the number of atoms in the in-plane direction, the intensities of the E$_\mathrm{2g}$ Raman mode should increase with the increasing number of layers of FGT. However, we did not observe any such intensity trend with thickness variation of FGT, which could be due to the weak intensity of E$_\mathrm{2g}^1$ mode. The reason behind weak intense in-plane vibrational Raman modes (E$_\mathrm{2g}$) is discussed below.\\
Our Raman spectrometer is set in back-scattering mode, where the laser is incident on the c-axis (+z direction) of the nanoflakes and the scattered light is collected by the same objective lens in the -z direction. The spectrometer is also set for the parallel geometry; $Z(XX)\bar{Z}$. Here, Z(XX)$\bar{Z}$ stands for the direction of the incident light (Z), polarization of incident light (X), polarization of scattered light (X), and direction of the scattered light ($\bar{Z}$). This geometry is mainly sensitive to the out-of-plane Raman modes (A$_\mathrm{1g}$). The in-plane Raman modes (E$_\mathrm{2g}$) can not be observed at all in this geometry, but if the polarization is not perfectly aligned in the X direction or the sample is not perfectly flat on the substrate, the (E$_\mathrm{2g}$) mode leakage with very weak intensities can be observed.\\
\begin{table*}
\caption{\label{tab:table1}The comparison of our experimentally observed Raman modes (cm$^{-1}$) of FGT with reported experimental and density functional theory (DFT) calculations.}
\begin{center}
\begin{tabular}{|c|c|c|c|c|}
\hline \hline \hline
Our results\, & Reported \cite{a8} & Reported \cite{a7}\, & Reported \cite{a7}\, & Reported \cite{a1}\\
(Experiment) & (Experiment) & (Experiment) & (DFT) & (DFT) \\ \hline \hline \hline
-- & -- & -- & E$_{2g}^1$: 28.4 & -- \\ \hline
-- & -- & -- & E$_{1g}^1$: 79.4 & -- \\ \hline
E$_{2g}^1$: 105 & E$_{2g}^1$: 108 & E$_{2g}^2$: 89.2 & E$_{2g}^2$: 115.5 & E$_{2g}^1$: 112 - 116 \\ \hline
A$_{1g}^*$: 129$^*$ & E$_{2g}^2$: 126 & -- & -- & -- \\ \hline
A$_{1g}^1$: 145 & A$_{1g}$: 144 &  A$_{1g}^1$: 121.1 & A$_{1g}^1$: 151.7 & A$_{1g}^1$: 140 - 146 \\ \hline
A$_{1g}^*$: 190$^*$ & -- & -- & -- & -- \\ \hline
E$_{2g}^2$: 230 & -- & -- & E$_{1g}^2$: 225.5 & E$_{2g}^2$: 224 - 231 \\ \hline
-- & -- & E$_{2g}^3$: 214 & E$_{2g}^3$: 238 & -- \\ \hline
A$_{1g}^2$: 280 & -- & A$_{1g}^2$: 239.6 & A$_{1g}^2$: 272 & A$_{1g}^2$: 275 - 280 \\ \hline
-- & -- & -- & E$_{2g}^4$: 362 & E$_{2g}^3$: 350 - 380 \\ \hline
\end{tabular}
\end{center}
\end{table*} 
Moreover, we observed three additional high-frequency Raman modes 190\,cm$^{-1}$, 230\,cm$^{-1}$ \& 280\,cm$^{-1}$ with low intensities, shown in Figure \ref{Raman modes} (a). Figure \ref{Raman modes} (b) is a zoomed image of the highlighted area of Figure \ref{Raman modes} (a). We observed these additional Raman modes, also in the encapsulated FGT shown in Figure \ref{Raman modes} (c) \& (d), which confirms the peaks belong to the intrinsic Raman modes. Xiangru Kong, \textit{et al.} reported the Raman active modes at 112 - 116\,cm$^{-1}$ (E$_{2g}^1$), 140 - 146\,cm$^{-1}$ (A$_{1g}^1$), 224 - 231\,cm$^{-1}$ (E$_{2g}^2$), 275 - 280\,cm$^{-1}$ A$_{1g}^2$) \& 350 - 356\,cm$^{-1}$ (E$_{2g}^3$) by First Principle calculations \cite{a1}. A Milosavljevi\'c, \textit{et al.} suggested eight Raman active modes, positioned at 28.4\,cm$^{-1}$ (E$_{2g}^1$), 79.4\,cm$^{-1}$ (E$_{1g}^1$), 115.5\,cm$^{-1}$ (E$_{2g}^2$), 151.7\,cm$^{-1}$ (A$_{1g}^1$), 225.5\,cm$^{-1}$ (E$_{1g}^2$), 238\,cm$^{-1}$ (E$_{2g}^3$), 272\,cm$^{-1}$ (A$_{1g}^2$) \& 362\,cm$^{-1}$ (E$_{2g}^4$) by their First Principle calculations and four modes, positioned at 89.2\,(E$_{2g}^2$), 121.1\,cm$^{-1}$ (A$_{1g}^1$), 214\,cm$^{-1}$ (E$_{2g}^3$) and 239.6\,cm$^{-1}$ (A$_{1g}^2$) by their experimental studies \cite{a7}. Ngoc-Toan Dang, \textit{et al.} observed three Raman modes experimentally, in FGT at 108\,cm$^{-1}$ (E$_{2g}^1$), 126\,cm$^{-1}$ (E$_{2g}^2$) and 144\,cm$^{-1}$ (A$_{1g}^1$) \cite{a8}. A comparative study of the above reports with our results is presented in Table \ref{tab:table1}.\\

The reported literature does not show the thickness-dependent Raman study experimentally; however, Xiangru Kong, \textit{et al.} reported the thickness-dependent Raman active modes using first-principles calculations. In our study, the Raman spectra were examined for various thicknesses (2-25 layers) of FGT. The spectra shows that the intensity of four Raman modes except two modes at 105\,cm$^{-1}$ and 230\,cm$^{-1}$, increases with the decrease in thickness of FGT nanofalkes, showing the behavior of out-of-plane vibrational mode (A$_{1g}$), therefore we assigned them as A$_{1g}$ peaks as shown in table \ref{tab:table1}. Both of the E$_{2g}$ modes are very weak Raman modes; therefore, we are unable to see any intensity dependence on the thickness of FGT. The E$_{2g}^2$: 230\,cm$^{-1}$ is studied by Luojun Du, \textit{et al.} and marked as a spin-phonon coupling mode \cite{a10}. Here, we carried out the Raman experiments at 300\,K at which the FGT possesses the paramagnetic phase as the ferromagnetic transition temperature is around 200\,K for FGT. Therefore, we observed the weak intensity of Raman mode at 230\,cm$^{-1}$ (E$_{2g}^2$). This mode does not show a systematic dependence on the thickness of FGT at room temperature. The A$_{1g}^2$: 280\,cm$^{-1}$ Raman mode shows the trivial behavior of out-of-plane phonon vibrational mode and is in good agreement with previous reports \cite{a7, a1}. In addition, we observe two additional Raman modes at 129\,cm $^{-1}$, cm$^{-1}$ \& 190\,cm$^{-1}$. The 129\,cm$^{-1}$ mode is observed by Ngoc-Toan Dang, \textit{et al.} \cite{a8} and it is marked as E$_{2g}^2$ mode, but in our results, we observed the thickness dependence which follows the trend of A$_{1g}$ mode. Another mode, at 190\,cm$^{-1}$, also shows the trend of the A$_{1g}$ mode with thickness variation; however, this mode is not observed earlier. Therefore, we marked the both modes; 129\,cm$^{-1}$ \& 190\,cm$^{-1}$, as A$_{1g}^*$. These two modes may be the consequences of the vibrations between adjacent layers of FGT, but the proper understanding remains unclear. As the thickness of FGT shows its signature in optical contrast and Raman studies, these two (optical contrast and Raman studies) can be utilized as a tool to determine the number of layers of FGT. Once the thickness of the FGT flake is measured by AFM and linked to the optical contrast values, which yield a linear relationship, one can easily identify the thickness of the FGT by the optical contrast and thickness plot without using AFM and without removing the sample from the glove box.

\section{Conclusion}
We report an efficient, non-destructive, and cost-effective technique for determining the number of layers of FGT. The technique utilizes optical contrast studies to determine the thickness of FGT flakes, which does not require additional equipment; an optical microscope is sufficient for measuring the thickness of FGT. It can be used inside the glove box; therefore, one does not need to extract the sample from the glove box, which prevents oxidation of FGT and allows for the study of the magnetic and transport behavior of pristine FGT of specific thickness. Additionally, we investigate the Raman spectra of FGT nanoflakes, which show the thickness dependence; however, the Raman spectroscopy alone can not determine the number of layers of FGT. we observed two additional Raman modes at 129\,cm$^-1$ \& 190\,cm$^-1$, that show the behavior of out-of-plane Raman mode with the thickness variation. These two modes may arise due to the vibrations between adjacent layers; however, the exact origin requires further investigation.

\section{Acknowledgments}
NY is thankful to the Council of Scientific \& Industrial Research (CSIR) for financial support. PD gratefully acknowledges the funding support from the Science \& Engineering Research Board (SERB) of Govt. of India (grant no. SPR/2021/000762). We acknowledge the Dept. of Physics, IIT Delhi for the Raman (funded by Department of Science and Technology, Govt. of India for Ultra Fast Optics facility through FIST program) and Atomic Force Microscope facilities. \\

\nocite{*}

\bibliography{Bib}

\end{document}